\documentclass[openacc]{rsproca_new}

\usepackage{cite}
\usepackage{appendix}
\usepackage[english]{babel}
\usepackage[utf8]{inputenc}
\usepackage[T1]{fontenc}
\usepackage{cancel,xcolor}
\usepackage{ulem}
\usepackage{caption}
\usepackage{dsfont}
\usepackage{amsmath}
\usepackage{amssymb}
\usepackage{amsfonts}
\usepackage{amsthm}
\usepackage{float}
\usepackage[final]{pdfpages}
\usepackage{graphicx}
\usepackage{hyperref}

\definecolor{dark-green}{RGB}{0, 128, 0}

\begin{document}

\title{Manifestation of Berry curvature in  geophysical ray tracing}

\author{N. Perez$^{1}$, P. Delplace$^{1}$ and A. Venaille$^{1}$}

\address{$^{1}$Univ Lyon, ENS de Lyon, Univ Claude Bernard, CNRS, Laboratoire de Physique (UMR CNRS 5672), F-69342 Lyon, France}

\subject{Geophysics, fluid mechanics, condensed matter physics}


\corres{Nicolas Perez\\
\email{nicolas.perez@ens-lyon.fr}}

\begin{abstract}

Geometrical phases, such as the Berry phase, have proven to be powerful concepts to understand numerous physical phenomena, from the precession of the Foucault pendulum to the quantum Hall effect and the existence of topological insulators. The Berry phase is generated by a quantity named Berry curvature, describing the local geometry of wave polarization relations and known to appear in the equations of motion of multi-component wave packets. Such a geometrical contribution in ray propagation of vectorial fields has been observed in condensed matter, optics and cold atoms physics. Here, we use a variational method with a vectorial Wentzel-Kramers-Brillouin (WKB) ansatz to derive ray tracing equations for geophysical waves and reveal the contribution of Berry curvature. We detail the case of shallow water wave packets and propose a new interpretation to their oscillating motion around the equator. Our result shows a mismatch with the textbook scalar approach for ray tracing, by predicting a larger eastward velocity for Poincaré wave packets. This work enlightens the role of wave polarization's geometry in various geophysical and astrophysical fluid waves, beyond the shallow water model.

\end{abstract}

\begin{fmtext}

\section{Introduction} \label{section:introduction}

The concept of geometric phase provides a unified framework to interpret physical phenomena as diverse as the Foucault pendulum experiment \cite{0} or the Aharonov-Bohm effect \cite{1}. It arises in virtually all fields of physics, including geophysical systems \cite{25,22,5b}. Those geometrical phases are the manifestation of an underlying geometrical property of the system, called the curvature $F$, such that the phase $\phi$ is formally related to it by an integral formula over a surface as $\phi = \frac{1}{2 \pi} \int F {\rm d} S$.

\end{fmtext}

\maketitle

For instance, for the Foucault pendulum, the geometrical phase corresponds to its precession angle, called Hannay angle, that results from Earth curvature. Here we are concerned with multi-component wave problems which generate a more abstract curvature, called Berry curvature. Such curvature encodes the polarization relations between the different wave components in parameter space. Although it was originally proposed in quantum physics \cite{1}, the emergence of a Berry curvature only requires the existence of several wave bands: each band $n$ is then characterized by a Berry curvature $F^{(n)}$. As such, it appears naturally in multi-component wave problems, from quantum condensed matter systems \cite{34,1b} to classical waves in optics \cite{1c,1d}, plasma physics, mechanics \cite{4c,4d} and hydrodynamics \cite{3,5}. A direct physical manifestation of the Berry curvature itself -- rather than the phase it is related to -- arises in ray tracing experiments: simply put, Berry curvature is a correction of the group velocity that bends rays trajectories \cite{34,7}. This correction was measured in condensed matter \cite{8,9}, cold atoms setups \cite{11} and photonic quantum walks \cite{10}. Although geophysical flows gather all the ingredients for a Berry curvature to emerge, it has up to now not been established how this curvature arises in geophysical ray tracing. This is the purpose of the present article.

Previous works have shown that inertia-gravity wave (or Poincaré wave) bands of the rotating shallow water model admit non-zero Berry curvature \cite{3}. This is therefore a natural starting point to address possible deviations of ray trajectories, that we derive using a variational principle inspired by quantum mechanics \cite{7,12}. More generally, we propose a consistent framework to reconsider the ray tracing problem in a large class of multi-component geophysical flow models, building upon previous results on vectorial elliptical linear systems \cite{13,14,14b}. This new approach emphasizes the role of Berry curvature effects that were up to now overlooked in geophysics.

To be more specific, we consider the dynamics of a wave packet in a medium characterized by a length $\ell$, varying spatially over a typical distance $L$. To fix the ideas, we will consider as main example shallow water wave packets evolving in a fluid layer. In that case $\ell$ is called the Rossby radius of deformation, which is inversely proportional to the rotation rate of the frame of reference \cite{15}.

The wave packet has a mean wavelength $\lambda$ and is modulated by an envelope of typical size $\Lambda > \lambda$. We assume that the extension of the wave's envelope $\Lambda$ is much smaller than the typical distance $L$ depicting the medium's variations ($\lambda, \Lambda \ll L$). This amounts to assuming that the solution is localized regarding the scale of the medium's variations, thus allowing one to project the packet onto a plane wave solution with wavelength $\lambda$ and local $\ell$ as parameters. We restrict ourselves to cases where the medium's properties enter the dispersion relation at lowest order, which, in the family of models considered in this paper, adds up to assuming $\ell \sim \lambda$. To sum up, the lengths describing the problem and its wave packet solution are organized as follows:
\begin{equation} \label{scales}
    \ell \sim \lambda < \Lambda \ll L \ .
\end{equation}

The usual way to derive ray tracing equations in geophysics is using a WKB ansatz \cite{16a,16,17} under scaling conditions such as \eqref{scales}. At the scale of the wavelength, the medium is approximately homogeneous and steady, which can be quantified by the introduction of a small dimensionless parameter, say $\varepsilon = \lambda / L$. This scale separation justifies the use of a WKB ansatz for a solution described by a scalar field $v$:
\begin{equation} \label{WKB}
    v(\mathbf{r},t) = {\rm e}^{{\rm i} \left(\frac{\Phi_0}{\varepsilon} + \Phi_1 + \mathcal{O}\left(\varepsilon\right) \right)} \ .
\end{equation}
Along with slowly varying (in time and space) functions $\Phi_i$ of order 1, expression \eqref{WKB} corresponds to the idea of a plane wave modulated by an envelope of finite size, with identifiable mean wave vector. In geophysics, a WKB ansatz is indeed generally injected in a scalar equation as if the problem were entirely defined by one scalar equation and a sufficient number of boundary conditions \cite{15,16,17}. Nevertheless, what about the models that are not well defined by and reducible to just one equation, even of higher order, and rather by a set of coupled wave equations with distinct coefficients ? Such a set of equations can be rewritten as one with a multi-component field, as will be illustrated in the following, which makes the WKB approximation more complex. Indeed a multi-component mode is characterized by both the dispersion and the polarization relations, the latter relating the different component fields between each other, all depending on the wave vector and the medium's parameters. Therefore, as the wave packet moves through the slowly varying medium, its polarization relations change accordingly as the ray coordinates $(\mathbf{r},\mathbf{k})$ evolve. As a consequence, we expect a footprint of the corresponding Berry curvature in the ray dynamics of multi-component geophysical wave problems.

The shallow water model with Coriolis parameter $f$ varying with meridional coordinate $y$ is a classical three-component wave model that cannot be reduced to one scalar equation. It will be used as the key example throughout this paper. A scalar equation for the meridional component of the velocity field $v(x,y,t)$ alone can be derived (see part \ref{part:comparison} of section \ref{section:application}), then one can inject ansatz \eqref{WKB} in this scalar equation in order to infer an approximate WKB solution for $v$ and find the corresponding relations for the phase functions at different orders. The results obtained via this approach for ray tracing can be found in various articles such as \cite{18,19,20} and textbooks \cite{15,21}, and are exposed as a reminder in part \ref{part:comparison} of section \ref{section:application}. However, the technique suffers from two difficulties:

\begin{itemize}
    \item First, the field $v$ alone is not sufficient to fully characterize the energy propagation. Indeed, the two other fields involved in the shallow water dynamics do not satisfy the same wave equation as $v$ when the Coriolis parameter is not homogeneous. Instead, they are both coupled to the field $v$. This leads to inconsistent WKB ansätze for the different fields of the problem, as noticed by Godin in the context of compressible stratified waves \cite{22}. In fact, keeping just a single scalar wave equation amounts to only considering the dispersion relation and loosing all the information on the polarization relations between the fields. In other words, the scalar approach misses any geometrical effect related to the polarization relations of the eigenmodes, which is an essential feature in the presence of inhomogeneity. Thereby the geometrical nature of the problem must be accounted for.
    \item Second and just as importantly, we find that the usual textbook derivation of ray tracing equations from a scalar equation involves a contradictory hypothesis on the scale of variation of the Coriolis parameter: keeping the variations of the Coriolis parameter at leading order of the WKB expansion (which is the dispersion relation that normally accounts for the homogeneous parameters) amounts to assuming that it varies considerably over a distance $\lambda$ (typically a few kilometers for geophysical shallow water waves on Earth), whereas it does over a scale $L$ that is much greater than $\lambda$ ($L$ is typically of a thousand kilometers on Earth). We show that it is therefore necessary to reconsider the order of magnitude of a small parameter characterizing the inhomogeneity of the medium in order to properly derive ray tracing equations at order one in this small parameter.
\end{itemize}

The issue of solving multi-component differential systems has long been addressed in applied mathematics \cite{23,24,13,14}. This has been related to the existence of geometric phases in the vectorial WKB solutions of multi-component wave problems \cite{25,26,14b,27,22}. However the role of Berry curvature in ray tracing -- which is a simpler problem because it does not aim at computing precisely the evolution of the phase -- has not been shown in the geophysical context. The purpose of the present work is to rectify the situation.

The paper is organized as follows. In section \ref{section:shallow-water} we introduce the shallow water model and its notations, the dimensionless parameters characterizing it and the spectral properties of the homogeneous problem, which are necessary to construct a wave packet solution for the slowly varying medium. Then in section \ref{section:Berry} we present a multi-component ansatz and a variational formulation of the problem, accounting for the different orders of the problem in a more consistent way and leading to the set of ray tracing equations. We show that the ray motion of inertia-gravity waves involves the Berry curvature and, as a byproduct, we report a direct analogy between the behaviour of inertia-gravity wave rays on $\beta$-plane Earth and the anomalous Hall effect \cite{28,34}. Finally in section \ref{section:application} we compare the result with the traditional scalar WKB approach for ray tracing.

\section{Formulation of the problem for shallow water waves} \label{section:shallow-water}

To illustrate the general issue of ray tracing for multi-component wave problems in geophysical setting, we propose to treat in this section the particular case of energy propagation through surface waves described by the shallow water model, keeping in mind that the results presented in section \ref{section:Berry} have a more general extent.

\subsection{The linearized rotating shallow water model} \label{part:model}

\begin{figure}[h!]
    \begin{center}
    \includegraphics[scale=0.35]{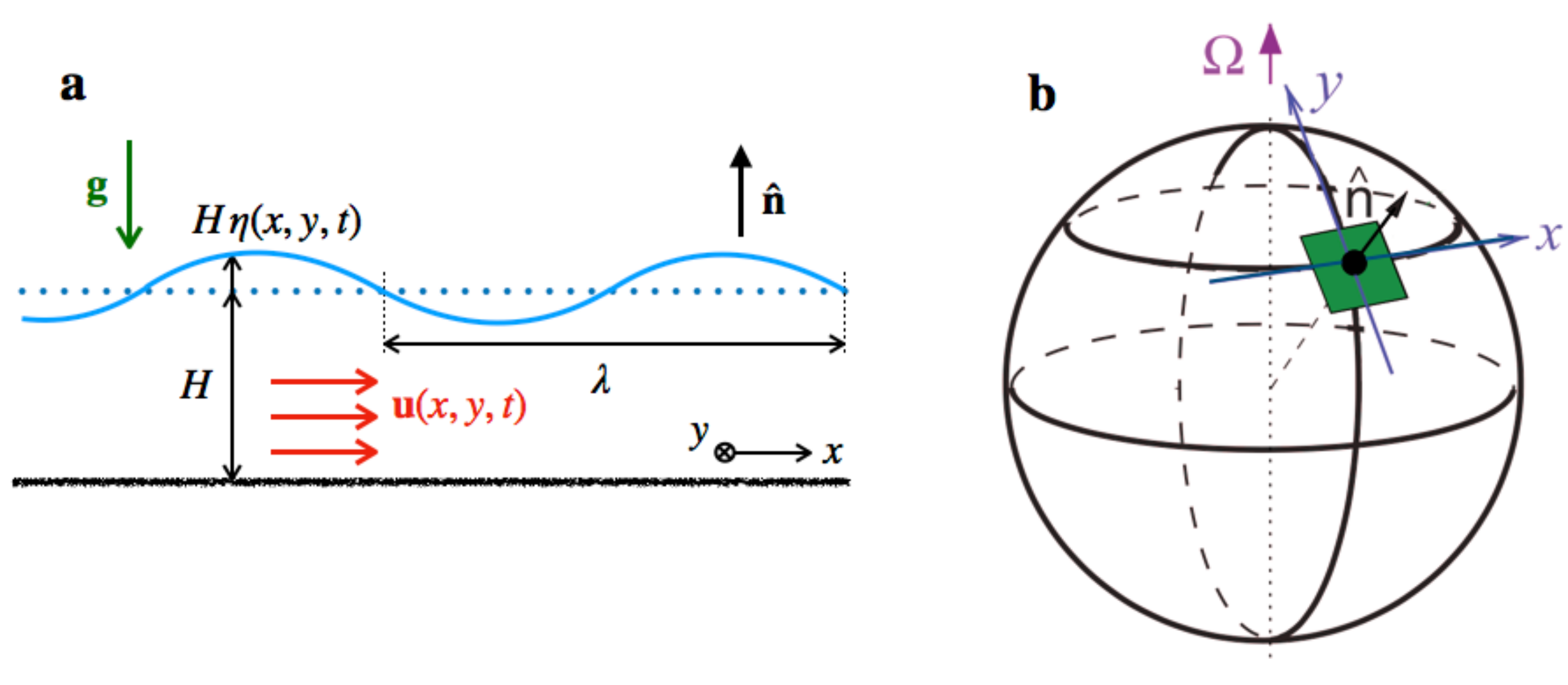}
    \end{center}
    \caption{\label{SW} ($\bold{a}$) The shallow water model describes wave propagation on the free surface of a thin layer of homogeneous fluid, whose total elevation is a function of time and position and reads $h \equiv H(1 + \eta)$. The mean depth $H$ must be much smaller than the typical horizontal variation scales, i.e. $H \ll \lambda$ with $\lambda$ the typical horizontal wavelength. In that case $c = \sqrt{gH}$. ($\bold{b}$) Planar approximation: the model is applied to a flow domain that takes place on a locally tangent plane on the planet, with local vertical basis vector $\hat{\mathbf{n}}$: with $\boldsymbol{\Omega}$ the rotation vector of Earth, the Coriolis parameter writes $f = 2 \boldsymbol{\Omega} \cdot \hat{\mathbf{n}}$.}
\end{figure}

Let us consider the shallow water model in planar geometry, describing a thin layer of fluid with homogeneous density in the presence of gravity (figure \ref{SW}). It is a bidimensional model, so the position reads $\mathbf{r} \equiv x \hat{\mathbf{e}}_{x} + y \hat{\mathbf{e}}_{y}$ and we write the planar velocity field $\mathbf{u} \equiv u \hat{\mathbf{e}}_{x} + v \hat{\mathbf{e}}_{y}$. For the sake of simplicity we focus on the case of a flat bottom, with constant phase speed denoted $c$ for surface waves. We consider a version of the model linearized around a state at rest $\mathbf{u} \equiv 0$ in the rotating frame of reference. We also use here the traditional approximation that consists in only considering the components of the Coriolis force in the horizontal plane, through the Coriolis parameter $f$, which is twice the projection of the planet's angular velocity vector on the local plane's vertical axis (see figure \ref{SW}). All fields, parameters and variables of the problem are nondimensionalized with length $\lambda$ and time $\lambda / c$. Then the shallow water equations governing the dynamics of the perturbation fields of velocity $u$, $v$ and elevation or geopotential $\eta$, all much smaller than 1 in amplitude, read
\begin{subequations} \label{shallow-water}
    \begin{align}
        \partial_t u &= - \partial_x \eta + f(y)v \label{eq_u} \ , \\
        \partial_t v &= - \partial_y \eta - f(y)u \label{eq_v} \ , \\
        \partial_t \eta &= - \partial_x u - \partial_y v \label{eq_eta} \ .
    \end{align}
\end{subequations}
Equations \eqref{eq_u} and \eqref{eq_v} are the projections of the linearized momentum conservation equation while \eqref{eq_eta} comes from the conservation of mass. Note that the only varying parameter in this problem is the adimensionalized Coriolis parameter $f$, which is central because it breaks the time-reversal symmetry of equations \eqref{shallow-water} and opens a gap in their dispersion relation (see figure \ref{courbure}). Equations \eqref{shallow-water} describe the free solutions of Laplace's tidal equations, in cartesian coordinates. The adimensionalized length
\begin{equation} \label{radius}
    \ell = \frac{1}{|f|}
\end{equation}
is the Rossby radius of deformation, and the gradient of $f$ with latitude is usually noted $\beta$.

\subsection{Dimensionless parameters of the problem and slow variables} \label{part:parameters}

We consider the problem of a shallow water wave packet evolving in a medium with varying Coriolis parameter. The dynamics is given by the linear rotating shallow water equations \eqref{shallow-water} and a typical initial condition as depicted by relations \eqref{scales}. The problem involves four length scales, and admits therefore three nondimensional parameters:
\begin{equation} \label{parameters}
    f, \quad \varepsilon \equiv \frac{\lambda}{L}, \quad \text{and} \quad \alpha \equiv \frac{\lambda}{\Lambda} \ .
\end{equation}
Let us comment on each of them:

\begin{itemize}
    \item $f \sim 1$ is a function of the slow variable $Y \equiv \varepsilon y$, varying over a typical scale $Y \sim 1$. We indiscriminately write $f$ as a function of $y$ or $Y$, as the case may be.
    \item $\varepsilon \ll 1$ is the inverse of the adimensionalized typical length for the Coriolis parameter's variation with latitude. It is the control parameter in the WKB expansion. The contribution of the Berry curvature to ray tracing equations arises as a correction of order $\varepsilon$, as will be shown in section \ref{section:Berry}.
    \item $\alpha \lesssim 1$ reflects the ratio between the wavelength and the spatial extension of the packet. For the wave packet to behave as a plane wave at leading order, its envelope's extension must be at least a few times the wavelength, and thus $\alpha$  must be smaller than one. The ratio $\varepsilon/\alpha = \Lambda/L$ appears as a small control parameter that guarantees that the medium is quasi-homogeneous at the scale of the wave packet. Therefore, $\alpha$ must not be too small (i.e. of the order of $\varepsilon$), otherwise this localized wave packet picture fails down.
\end{itemize}
   
We propose to organize these parameters according to the set of inequalities $\varepsilon \ll \alpha^2 \lesssim 1$ to derive ray tracing equations, together with $f(Y) \sim 1$. This last choice guarantees that rotation enters at lowest order in the wave dispersion relation. The reason for the scaling $\varepsilon \ll \alpha^2$ will be clarified in part \ref{part:variational1} of section \ref{section:Berry}.

Consequently, we suppose that the phase functions are natural functions of the slow variables
\begin{equation} \label{slow1}
    (X,Y,T) \equiv (\mathbf{R},T) \equiv (\varepsilon x, \varepsilon y, \varepsilon t) \ ,
\end{equation}
whereas the amplitude function is by definition a natural function of the variables
\begin{equation} \label{slow2}
    (X',Y',T') \equiv (\mathbf{R}',T') \equiv (\alpha x, \alpha y, \alpha t) \ .
\end{equation}
Therefore the existence of these two parameters describing the solution leads to a multiscale characterization of the ansatz. In the following we sometimes rather use the derivation operators with respect to these coordinates, namely
\begin{equation} \label{slow_nabla}
    \nabla_\mathbf{R} \equiv (\partial_X, \partial_Y) \text{ and }\nabla_{\mathbf{R}'} \equiv (\partial_{X'}, \partial_{Y'}) \ .
\end{equation}
Introducing these notations does not mean that we aim at computing the evolution of the wave packet's phase, which is not relevant in this ray tracing context. Our assumptions, however, allow us to treat the variation of the Coriolis parameter as a perturbation -- as it should be, considering its order of magnitude -- that will enter ray tracing equations at lower order. As explained in the introduction, we need such geometric tools as the Berry curvature and a vectorial WKB ansatz in order to address this perturbation correctly regarding the multi-component nature of shallow water waves.

\subsection{Spectral and geometrical features of waves in the $f$-plane model} \label{part:spectral}

Following \cite{17}, in order to construct the wave packet solution for waves in a slowly varying medium we must first and foremost derive the plane wave solutions supported by the translationally invariant medium. Equations \eqref{shallow-water} can be rewritten in a more compact Schrödinger-like equation
\begin{align}
    &\left( {\rm i} \partial_t - \hat{\mathcal{H}}(\mathbf{r},\boldsymbol{\nabla}) \right) \psi(\mathbf{r},t) = 0 \label{schrödinger} \ , \\
    \text{with} \quad &\hat{\mathcal{H}}(\mathbf{r},\boldsymbol{\nabla}) \equiv \begin{pmatrix} 0 & {\rm i} f(y) & -{\rm i} \partial_x \\ -{\rm i} f(y) & 0 & -{\rm i} \partial_y \\ -{\rm i} \partial_x & -{\rm i} \partial_y & 0 \\ \end{pmatrix} \ , \qquad \psi(\mathbf{r},t) \equiv \begin{pmatrix} u \\ v \\ \eta \\ \end{pmatrix} \label{operator} \ ,
\end{align}
where $\hat{\mathcal{H}}(\mathbf{r},\boldsymbol{\nabla})$ is a linear differential operator and $\psi(\mathbf{r},t)$ a three-components vector field, called the \textit{wave function} hereafter.

\begin{figure}[h!]
    \centering
    \includegraphics[scale=0.35]{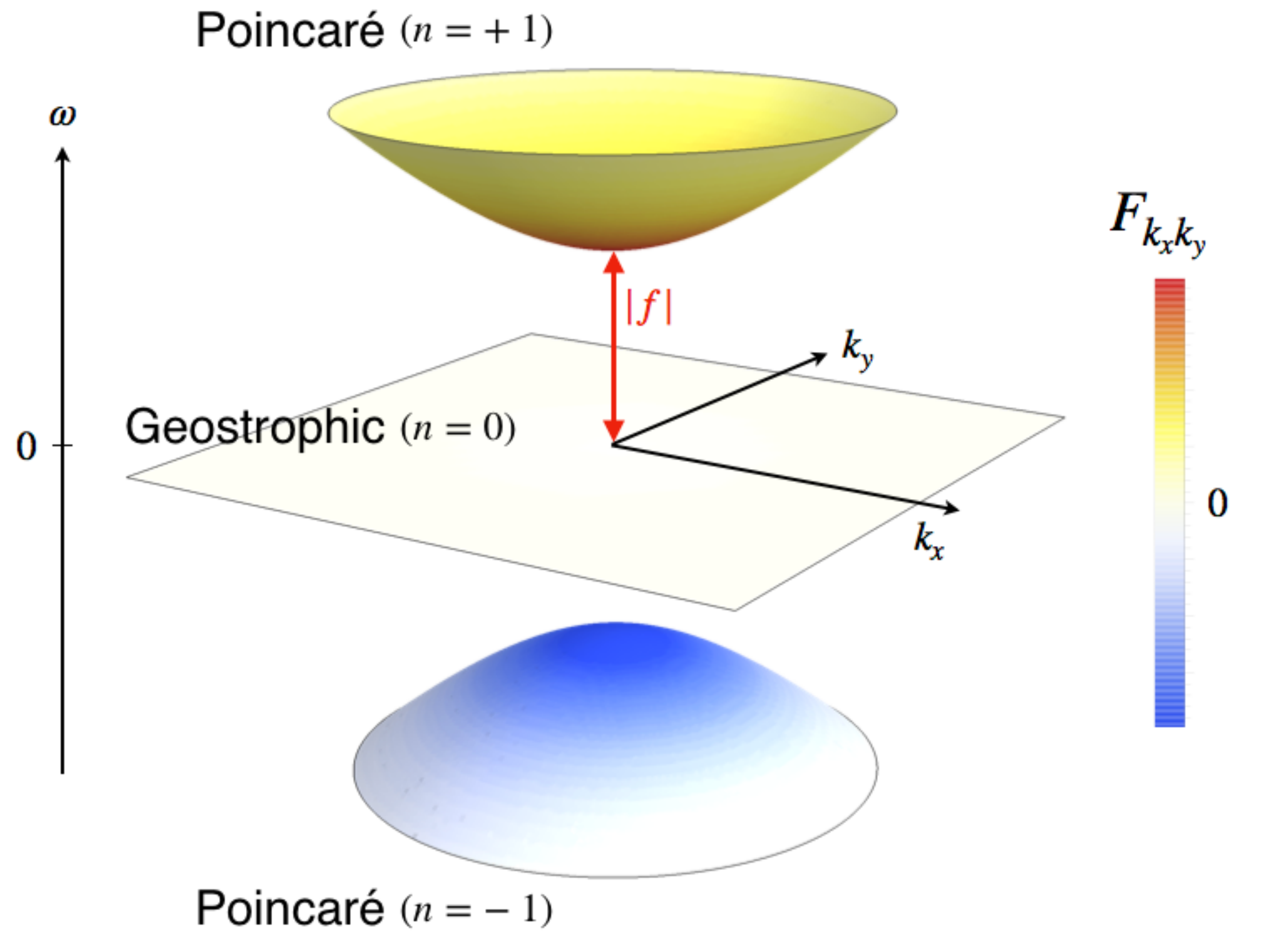}
    \caption{\label{courbure} Plot of the dispersion relation of $f$-plane shallow water equations for the three bands (indexed by n = -1, 0, +1). The gap between them equals $|f|$. In color is the value of the component $F_{k_{x} k_{y}}$ of the Berry curvature, defined in equation \eqref{curvature}, as a function of $\mathbf{k}$ and for a fixed positive value of $f$, for each band.}
\end{figure}

It is instructive to consider the simplest possible case where the Coriolis parameter does not vary in space, the celebrated $f$-plane approximation (figure \ref{SW}), originally studied by Kelvin \cite{29}. Plane wave solutions are then allowed and written in the form $\psi(\mathbf{r},t) = {\rm e}^{{\rm i}(\mathbf{k} \cdot \mathbf{r} - \omega t)} U_{\mathbf{k},\omega}$. The problem \eqref{schrödinger} boils down to a matricial equation
\begin{equation} \label{symbol}
    H[f,\mathbf{k}] U_{\mathbf{k},\omega} = \omega U_{\mathbf{k},\omega} \ , \quad \text{with} \quad H[f,\mathbf{k}] \equiv \begin{pmatrix} 0 & {\rm i} f & k_x \\ -{\rm i} f & 0 & k_y \\ k_x & k_y & 0 \\ \end{pmatrix} ,
\end{equation}
where $U_{\mathbf{k},\omega} \equiv \begin{pmatrix} \hat{u}_{\mathbf{k},\omega} & \hat{v}_{\mathbf{k},\omega} & \hat{\eta}_{\mathbf{k},\omega} \end{pmatrix}^t \in \mathbb{C}^3$ is the vector of Fourier transforms and $\mathbf{k} \equiv (k_x, k_y)$. Normalized eigenmodes and eigenvalues of this problem are denoted respectively $\omega_{n}[f,\mathbf{k}]$ and $U_{n}[f,\mathbf{k}]$, with $n \in \{-1,0,1\}$ for the shallow water model. In this particular case one gets
\begin{equation} \label{spectrum}
\omega_0 = 0 \ , \quad \omega_{\pm 1} = \pm \sqrt{f^2 + \mathbf{k}^2} \ .
\end{equation}
The first one is known as geostrophic modes and the seconds as inertia-gravity or Poincaré waves. The dispersion relation is depicted in figure \ref{courbure}. Polarization relations of the waves are given by the eigenmodes $U_{n}[f,\mathbf{k}]$ provided in appendix \ref{appendix:eigenv}.

The normalized eigenmodes $U_{n}$ are parameterized over a base space $(f,k_x, k_y)$. At each point of this base space, each vector $U_{n}$ is defined up to a phase, which defines a complex eigenspace of dimension 1. The continuous family of such parametrized eigenspaces over $(f,k_x, k_y)$ defines a \textit{fiber bundle}, or \textit{eigenbundle}, which owns specific geometric properties, whose repercutions on wave packet dynamics is the central motivation of this work. Those properties can be quantified by the Berry curvature, defined as
\begin{equation} \label{curvature}
    F^{(n)}_{\lambda_\mu \lambda_\nu} = {\rm i} \left( \frac{\partial U_{n}^\dagger}{\partial \lambda_\mu}\frac{\partial U_{n}}{\partial \lambda_\nu} - \frac{\partial U_{n}^\dagger}{\partial \lambda_\nu}\frac{\partial U_{n}}{\partial \lambda_\mu} \right) \ ,
\end{equation}
where the $\lambda_\mu$ are the coordinates of the base space, e.g. $\boldsymbol{\lambda} \equiv (f,k_x, k_y)$ here. Importantly, this quantity is a gauge-invariant real-valued quantity: it does not depend on the phase choice for the eigenmodes $U_{n}$. Qualitatively, it describes local twisting of these eigenmodes. It was computed for shallow water waves in $f$-plane, i.e. for a fixed value of $f$, in \cite{3}. The value of $F_{k_x k_y}$ for $f$-plane waves is reproduced on the dispersion relation in figure \ref{courbure}.

Let us now turn into a more general case of rotating shallow water waves where $f(\mathbf{r})$ varies spatially. It is useful to consider the \textit{symbol} $H[f(\mathbf{r}),\mathbf{k}]$, which in our case is formally defined by replacing $\boldsymbol{\nabla}$ in the operator $\hat{\mathcal{H}}(\mathbf{r},\boldsymbol{\nabla})$ of \eqref{operator} by ${\rm i} \mathbf{k}$, as in equation \eqref{symbol} for shallow water equations. Eigenmodes and eigenvalues of the symbol are just those of the $f$-plane problem. The study of symbols is at the heart of a mathematical framework called Weyl calculus \cite{23}. It provides a powerful tool to study wave problems in physics, starting from semi-classical analysis, see e.g. \cite{13,14,32}. The strong interest of this approach in geophysical wave transport has recently been advocated by Onuki \cite{14b}. We show in section \ref{section:Berry} how the Berry curvature emerges in this framework.

\section{Berry curvature in ray tracing of multi-component waves} \label{section:Berry}

In this section we derive ray tracing equations by exploiting the equivalence between the Schrödinger-like equation \eqref{schrödinger} and a problem of functional optimization \cite{12,7}. First we introduce a vectorial ansatz for the wave packet, then we show that the Lagrangian describing its dynamics involves the Berry connection of the wave's eigenspace. The results of this section are not specific to the shallow water model, and can be applied to any other continuous multi-component wave system provided that the scaling discussed in part \ref{part:parameters} of section \ref{section:shallow-water} is respected.

\subsection{A ray tracing method based on a variational principle for multi-component wave problems} \label{part:variational1}

Our starting point to address the Schrödinger-like equation \eqref{schrödinger} is the following vectorial ansatz:
\begin{equation} \label{space-ansatz}
    \psi'(\mathbf{r},t) = A(\alpha \mathbf{r}, \alpha t) \, {\rm e}^{{\rm i} \Phi (\varepsilon \mathbf{r}, \varepsilon t)} \, U_{n} [f(\varepsilon y),\mathbf{k}_{c} (\varepsilon t)] \ .
\end{equation}

This ansatz corresponds to the initial condition discussed in introduction, which is a wave packet on band $n$ with adiabatically varying mean wave vector $\mathbf{k}_{c}(T)$ and corresponding local eigenmode $U_{n} [f(Y),\mathbf{k}_{c}(T)]$, modulated by the envelope function $A(X',Y',T')$ (with the notations introduced in relations \eqref{slow1} and \eqref{slow2} for the variables $X,Y,T,X',Y'$ and $T'$). Again, all functions in ansatz \eqref{space-ansatz} can be written as a function of the slow or normal variables in the following, as the case may be. Both the slowly varying phase $\Phi(X,Y,T)$ and the envelope function $A$ of the vectorial field are real-valued, contrary to the scalar ansatz \eqref{WKB}, where the phase was given by $\mathcal{R}e \Phi$ and the amplitude by $\exp{\left( -\mathcal{I}m \Phi \right)}$, which is more convenient for the scalar method. We can expand $\Phi = \Phi_0 / \varepsilon + \Phi_1 + ...$, with all $\Phi_i$ of order 1, in a WKB framework, but we do not expand further than $i=1$, all the geometrical effect at interest here being expected to appear at this order. One needs to eventually take the real part of the wave function \eqref{space-ansatz} to find the actual solution of the fluid problem. It is however more convenient to work with the complex solutions throughout this paper.

We aim at deriving the ray tracing equations, meaning the equations of motion of the coordinates defining the wave packet: the position $\mathbf{r}_{c}(t)$ and momentum $\mathbf{k}_{c}(t)$ of its center of mass, that we hereby define. It is convenient to use a bra-ket notation, although with non-normalized vectors. For instance, by definition of the canonical inner product of Hilbert space $L_2 (\mathbb{R}^2,\mathbb{C}^3)$:
\begin{equation} \label{product}
    \langle \psi(t) | \psi(t) \rangle \equiv \iint {\rm d} x {\rm d} y \, \psi(x,y,t)^\dagger \psi(x,y,t) = \int {\rm d}^2 r \, A(\alpha \mathbf{r}, \alpha t)^2 \ ,
\end{equation}
the adimensionalized mechanical energy of the perturbation reads
\begin{equation} \label{energy}
    E = \frac{1}{2} \iint {\rm d} x {\rm d} y \left( |u(x,y,t)|^2 + |v(x,y,t)|^2 + |\eta(x,y,t)|^2 \right) = \frac{1}{2} \langle \psi | \psi \rangle \ .
\end{equation}
The energy is a constant of motion since the operator $\hat{\mathcal{H}}$ is Hermitian. Similarly the position and momentum of center of the wave packet are defined with Hermitian operators as
\begin{equation} \label{position-momentum}
    \mathbf{r}_{c} \equiv \frac{\langle \psi | \mathbf{r} | \psi \rangle}{\langle \psi | \psi \rangle} \ , \quad \text{and} \quad \mathbf{k}_{c} \equiv \frac{\langle \psi | (-{\rm i} \boldsymbol{\nabla}) | \psi \rangle}{\langle \psi | \psi \rangle} \ .
\end{equation}
The latter must be equal to the wave vector preassigned in ansatz \eqref{space-ansatz}. This consistency condition yields the following relation, shown in appendix \ref{appendix:spacevaria}:
\begin{equation} \label{k=K}
    \mathbf{k}_{c}(t) = \nabla_\mathbf{R} \Phi_0 + \varepsilon \left( \nabla_\mathbf{R} \Phi_1 - {\rm i} U_{n}^\dagger \nabla_\mathbf{R} U_{n} \right) + \mathcal{O}\left(\left(\frac{\varepsilon}{\alpha}\right)^2\right) \ ,
\end{equation}
all the phase functions in relation \eqref{k=K} being evaluated at time $t$ and position $\mathbf{r}_{c}(t)$, eigenmode $U_n$ at $[f(Y_{c}(T)),\mathbf{k}_{c}(T)]$.

Solving equation \eqref{schrödinger} is equivalent to the variational problem of finding the function $\psi' (\mathbf{r}, t)$ optimizing the action integrated from the following Lagrangian \cite{12,7}:
\begin{equation} \label{Lagrangian}
    \mathcal{L'}[\psi'] \equiv \frac{\langle \psi' | \left( {\rm i}\partial_t - \hat{\mathcal{H}}(\mathbf{r},\boldsymbol{\nabla}) \right) | \psi' \rangle}{\langle \psi' | \psi' \rangle} \ .
\end{equation}
The solution of the variational problem \eqref{Lagrangian} directly gives a solution of equation \eqref{schrödinger}, differing by a time-dependent phase that does not change any of the observables defined by relations \eqref{energy} and \eqref{position-momentum} \cite{12}.

This variational approach is convenient for ray tracing because the Lagrangian $\mathcal{L'}$ defined by expression \eqref{Lagrangian} depends only on the variables of center of mass $\mathbf{r}_{c}$ and $\mathbf{k}_{c}$ and their time derivatives at lowest orders (see appendix \ref{appendix:spacevaria} for the detailed derivation):
\begin{equation} \label{Lagrangian_red}
    \mathcal{L'} = \dot{\mathbf{r}}_{c} \cdot \mathbf{k}_{c} + \dot{\mathbf{r}}_{c} \cdot {\rm i} U_{n}^\dagger \frac{\partial U_{n}}{\partial \mathbf{r}_{c}} + \dot{\mathbf{k}}_{c} \cdot {\rm i} U_{n}^\dagger \frac{\partial U_{n}}{\partial \mathbf{k}_{c}} - \Omega^{(n)} (\mathbf{r}_{c},\mathbf{k}_{c}) - \frac{{\rm d}}{{\rm d} t} \chi(t) + \mathcal{O}\left(\left(\frac{\varepsilon}{\alpha}\right)^2\right) \ ,
\end{equation}
with the notations $\dot{X} \equiv {\rm d} X / {\rm d} t$ and
\begin{equation} \label{ray_Hamiltonian}
    U_{n} (\mathbf{r}_{c},\mathbf{k}_{c}) \equiv U_{n} [f(\mathbf{r}_{c}),\mathbf{k}_{c}] \ , \quad \text{and} \quad \Omega^{(n)} (\mathbf{r}_{c},\mathbf{k}_{c}) \equiv \frac{\langle \psi | \hat{\mathcal{H}} | \psi \rangle}{\langle \psi | \psi \rangle} + \mathcal{O}\left(\left(\frac{\varepsilon}{\alpha}\right)^2\right) \ ,
\end{equation}
where $| \psi \rangle$ is the wave function and index $n$ indicates the band over which the packet is defined, as written in expression \eqref{space-ansatz}. In other words, defining the reduced Lagrangian as
\begin{equation} \label{Lagrangian_bis}
    \mathcal{L}(\mathbf{r}_{c},\mathbf{k}_{c},\dot{\mathbf{r}}_{c},\dot{\mathbf{k}}_{c}) \equiv \dot{\mathbf{r}}_{c} \cdot \mathbf{k}_{c} + \dot{\mathbf{r}}_{c} \cdot {\rm i} U_{n}^\dagger \frac{\partial U_{n}}{\partial \mathbf{r}_{c}} + \dot{\mathbf{k}}_{c} \cdot {\rm i} U_{n}^\dagger \frac{\partial U_{n}}{\partial \mathbf{k}_{c}} - \Omega^{(n)} (\mathbf{r}_{c},\mathbf{k}_{c}) \ ,
\end{equation}
the difference between functionals $\mathcal{L'}$ and $\mathcal{L}$ adds up to second order terms and the exact time derivative of a function $\chi(t)$ that can be removed, for it does not affect the solutions of the variational problem (see appendix \ref{appendix:spacevaria}).

We recognize in expression \eqref{Lagrangian_bis} for $\mathcal{L}$ the Berry connection ${\rm i} U_{n}^\dagger {\rm d} U_{n}$ of the $n^{th}$ eigenmodes' bundle parametrized over the ray phase space $(\mathbf{r}_{c},\mathbf{k}_{c})$ as base space (cf. part \ref{part:spectral} of section \ref{section:shallow-water}), with
\begin{equation} \label{connection_time}
    \dot{\mathbf{r}}_{c} \cdot {\rm i} U_{n}^\dagger \frac{\partial U_{n}}{\partial \mathbf{r}_{c}} + \dot{\mathbf{k}}_{c} \cdot {\rm i} U_{n}^\dagger \frac{\partial U_{n}}{\partial \mathbf{k}_{c}} = {\rm i} U_{n}^\dagger \frac{{\rm d} U_{n}}{{\rm d} t} \ .
\end{equation}
We explain in part \ref{part:variational2} of this section the relation between this function and the Berry curvature introduced in section \ref{section:shallow-water}. This quantity connects $U_{n} (\mathbf{r}_{c} + {\rm d} \mathbf{r}_{c},\mathbf{k}_{c} + {\rm d} \mathbf{k}_{c})$ to $U_{n} (\mathbf{r}_{c},\mathbf{k}_{c})$, thus providing a footprint of the vectorial polarization relations in the ray trajectory in phase space. If the connection were null we would recover the simple form $\mathcal{L} = \dot{\mathbf{r}}_{c} \cdot \mathbf{k}_{c} - \Omega^{(n)} (\mathbf{r}_{c},\mathbf{k}_{c})$ and simply identify observables $\mathbf{r}_{c}$, $\mathbf{k}_{c}$ and $\Omega^{(n)}$ - defined in equations \eqref{position-momentum} and \eqref{ray_Hamiltonian} - to the generalized position, momentum\footnote{Indeed $\mathbf{r}_{c}$ and $\mathbf{k}_{c}$ are both seen as generalized coordinates in phase space, but in the particular case of a null connection we have the equality $\mathbf{k}_{c} = \partial \mathcal{L}/\partial \dot{\mathbf{r}}_{c}$, therefore $\mathbf{k}_{c}$ identifies to the conjugated momentum of $\mathbf{r}_{c}$.} and Hamiltonian\footnote{This corresponds to the ray Hamiltonian, which has the dimension of a frequency and is not to be confused with the energy Hamiltonian commonly used in variational methods for fluid dynamics \cite{17,30,41}.} of classical analytical mechanics, respectively. Such a trivial case would occur for instance in a one-band system or, more generally, in the case of a band with topologically flat eigenbundles: all Berry connections, of the form ${\rm i} U_{n}^\dagger \partial U_{n} / \partial q$ as appearing in expression \eqref{connection_time}, are null in that situation.

Expression \eqref{Lagrangian_bis} justifies the interest of this variational principle altogether with the approximation of a wave packet in a slowly varying medium: the dynamics of such solution can be reduced to that of its coordinates of center of mass. The results of this section do not depend on the choice of ansatz \eqref{space-ansatz}, as long as the scaling $\varepsilon \ll \alpha^2 \lesssim 1$, introduced in part \ref{part:parameters} of section \ref{section:shallow-water}, is respected. More precisely, the reduced Lagrangian $\mathcal{L}$ defined in equation \eqref{Lagrangian_bis} and the full Lagrangian $\mathcal{L}^{\prime}$ defined in equation \eqref{Lagrangian} are equal up to order $\varepsilon$ in the regime  $(\varepsilon/\alpha)^2 \ll \varepsilon \ll 1 $.

\subsection{Ray tracing equations and Berry curvature} \label{part:variational2}

The coordinates of the wave packet in phase space, $\mathbf{r}_{c}(t)$ and $\mathbf{k}_{c}(t)$, are the generalized coordinates of the system whose Lagrangian $\mathcal{L}$ is defined by equation \eqref{Lagrangian_bis}, and their time derivatives are the generalized velocities. Cancelling the lowest-order variations of the action integrated from $\mathcal{L}$ yields the Euler-Lagrange equations:
\begin{equation} \label{Euler-Lagrange}
    \frac{\partial \mathcal{L}}{\partial q} = \frac{{\rm d}}{{\rm d} t}\left(\frac{\partial \mathcal{L}}{\partial \dot{q}}\right) \ ,
\end{equation}
where $q$ is any of $\mathbf{r}_c$ or $\mathbf{k}_c$ components. Coordinates $x$ and $y$ will be noted with Greek letters $\mu, \nu$ for Einstein tensorial notations in the following. From the Berry connection arising in expression \eqref{Lagrangian_bis} results that the conjugated momentum of $\mathbf{r}_c$ is equal to $\mathbf{k}_c$ plus an additional geometric term related to the Berry connection:
\begin{equation} \label{generalized_momentum}
    \frac{\partial \mathcal{L}}{\partial \dot{\mathbf{r}}_{c}} = \mathbf{k}_{c} + {\rm i} U_{n}^\dagger \frac{\partial U_{n}}{\partial\mathbf{r}_{c}} \ , \quad \text{and} \quad \frac{\partial \mathcal{L}}{\partial \dot{\mathbf{k}}_{c}} = {\rm i} U_{n}^\dagger \frac{\partial U_{n}}{\partial\mathbf{k}_{c}}
\end{equation}
for the conjugated momentum of $\mathbf{k}_c$. Deriving expression \eqref{Lagrangian_bis} and injecting expressions \eqref{generalized_momentum} in the Euler-Lagrange equations \eqref{Euler-Lagrange} finally yields, with $q = \mathbf{k}_c$ and $q = \mathbf{r}_c$ respectively:
\begin{subequations} \label{ray_tracing}
    \begin{align}
        \dot{r}_{\mu c} &= +\frac{\partial \Omega^{(n)}}{\partial k_{\mu c}} - F^{(n)}_{k_{\mu c} r_{\nu c}} \dot{r}_{\nu c} - F^{(n)}_{k_{\mu c} k_{\nu c}} \dot{k}_{\nu c} \label{ray_tracing_position} \ , \\
        \dot{k}_{\mu c} &= -\frac{\partial \Omega^{(n)}}{\partial r_{\mu c}} + F^{(n)}_{r_{\mu c} r_{\nu c}} \dot{r}_{\nu c} + F^{(n)}_{r_{\mu c} k_{\nu c}} \dot{k}_{\nu c} \label{ray_tracing_vector} \ .
    \end{align}
\end{subequations}
The Berry curvature tensor $F^{(n)}$ is defined as in equation \eqref{curvature}, with $\lambda_\mu \in \{ x_c, y_c, k_{xc}, k_{yc} \}$ for the shallow water model, choosing the phase space coordinates to parametrize the eigenmodes $U_{n}(\boldsymbol{\lambda}) \equiv U_{n}(\mathbf{r}_{c},\mathbf{k}_{c})$ of the symbol $H(\boldsymbol{\lambda}) \equiv H[f(\mathbf{r}_{c}),\mathbf{k}_{c}]$, with eigenvalue $\omega_{n}(\mathbf{r}_{c},\mathbf{k}_{c}) \equiv \omega_{n}[f(\mathbf{r}_{c}),\mathbf{k}_{c}]$. Since $U_{n}$ is normalized for every $(\mathbf{r}_{c},\mathbf{k}_{c})$, the Berry connection ${\rm i} U_{n}^\dagger {\rm d} U_{n}$ has real coefficients.

As shown in \cite{1}, the Berry curvature of the $n^{th}$ wave bundle defined in equation \eqref{curvature} can be expressed with off-diagonal elements of the symbol's derivatives:
\begin{equation} \label{curvature_bis}
    F^{(n)}_{\lambda_\mu \lambda_\nu} = -2 \, \mathcal{I}m \left( \sum_{m \neq n} \frac{U_{n}^{\dagger} \frac{\partial H}{\partial \lambda_\mu} U_{m} \, U_{m}^{\dagger} \frac{\partial H}{\partial \lambda_\nu} U_{n}}{(\omega_{n} - \omega_{m})^2} \right) \quad \text{(Berry curvature)} \ .
\end{equation}
As for $\Omega^{(n)}$, defined by relation \eqref{ray_Hamiltonian}, we show the following expression in appendix \ref{appendix:spacevaria}:
\begin{equation} \label{correction}
    \Omega^{(n)} - \omega_{n} = - \mathcal{I}m \left( \sum_{m \neq n} \frac{U_{n}^{\dagger} \frac{\partial H}{\partial r_{\mu c}} U_{m} \, U_{m}^{\dagger} \frac{\partial H}{\partial k_{\mu c}} U_{n}}{\omega_{n} - \omega_{m}} \right) \quad \text{(gradient correction)} \ .
\end{equation}
The \textit{gradient correction} $\Omega^{(n)} - \omega_n$ comes from the spatial variations of the medium's parameters: it is a \textit{dynamical} quantity. Conversely, the curvature $F^{(n)}$ as expressed in expressions \eqref{curvature} or \eqref{curvature_bis} only depends on the choice made to parametrize the manifold over which the eigenmodes are defined: it is a purely \textit{geometric} quantity. As a by-product, $\Omega^{(n)} - \omega_n$ is exactly null for the homogeneous problem, contrary to the Berry curvature $F^{(n)}$. Both the ray Hamiltonian and the Berry curvature corrections in \eqref{ray_tracing} are of order $\varepsilon$ at best, therefore we recover the classical scalar ray tracing equations $\dot{\mathbf{r}}_{c} = \partial \omega_{n} / \partial \mathbf{k}_{c}$ and $\dot{\mathbf{k}}_{c} = -\partial \omega_{n} / \partial \mathbf{r}_{c}$ at leading order, with $\omega_n$ given by the dispersion relation.

Expressions \eqref{curvature_bis} and \eqref{correction} give respectively the curvature and the ray Hamiltonian with interband terms $m \neq n$, showing that the corrections are larger at points $\mathbf{k}$ where the band $n$ is close to another one in frequency and vanish as the bands are separated in frequency\footnote{The same perturbative expressions for equations \eqref{curvature_bis} and \eqref{correction} can be found in \cite{1,7} for corrected Bloch bands, due to the interaction between an external magnetic field and the orbital magnetic moment of an electron, for instance.}. This effect is visible in figure \ref{courbure} in the shallow water case, with higher values of $F_{k_{x} k_{y}}$ when the triplet $(f,k_{x},k_{y})$ becomes closer to the degeneracy point $(0,0,0)$. We shall keep in mind that $\omega_n$, $\omega_m$, $\Omega
^{(n)}$, $H$, $U_n$ and $U_m$ in these relations are all functions of $(\mathbf{r}_{c},\mathbf{k}_{c}) = (x_c, y_c, k_{xc}, k_{yc})$.

The set of equations \eqref{ray_tracing} is different from the ones obtained with the traditional scalar derivation (see part \ref{part:comparison} of section \ref{section:application}) as it accounts for both the corrected ray Hamiltonian $\Omega^{(n)}$ and the geometry of the wave packet's polarization relations in ray phase space through the Berry curvature. Such a form of ray tracing equations has already been derived for different purposes such as the study of the time-dependent Schrödinger equation in the adiabatic limit \cite{12}, the theory of Bohr-Sommerfeld quantization \cite{32} and semiclassical analyses of wave packet trajectories in slowly perturbed crystals \cite{7,33,34,35}, optical lattices \cite{10,36} and ultracold atoms \cite{11}. Formal expressions for first order corrections to transport equations in multi-component fluid wave problems with inhomogeneous media were recently proposed by Onuki, with concrete applications to shallow water waves \cite{14b}. However, the explicit contribution of Berry curvature to ray tracing equations has never been exhibited in astrophysical or geophysical fluid dynamics. Since non-trivial Berry curvature has recently been reported in magnetized plasma \cite{4} and compressible stratified fluids \cite{5}, we expect that geometrical corrections to ray tracing equations will matter in those context. For now, we illustrate this effect by focusing on shallow water wave rays with varying backgroung rotation.

\section{Application to shallow water wave packets} \label{section:application}

\subsection{$\beta$-plane shallow water waves} \label{part:observables}

As shown in section \ref{section:Berry}, the ray equations of motions \eqref{ray_tracing} require only the knowledge of the ray Hamiltonian $\Omega^{(n)} (\mathbf{r}_{c},\mathbf{k}_{c})$ and the Berry curvature tensor $F^{(n)} (\mathbf{r}_{c},\mathbf{k}_{c})$, which is a 4 by 4 matrix in the particular case of shallow water equations. In other words, with equations \eqref{ray_tracing} we are dealing with a double correction to the usual ray tracing equations. These additional terms appear at lowest order in the medium's variations, since their expressions only involve space derivatives of the symbol $H[f(\mathbf{r}_{c}),\mathbf{k}_{c}]$ and its eigenmodes. We will therefore only need gradient corrections of the Coriolis parameter to apply equations \eqref{ray_tracing} to shallow water wave rays. It is therefore useful to introduce a new quantity:
\begin{equation} \label{beta}
    \beta(y) \equiv \frac{{\rm d} f(y)}{{\rm d} y} = \varepsilon \frac{{\rm d} f(Y)}{{\rm d} Y} \ .
\end{equation}
By definition $\beta$ is of order $\varepsilon$. When $\beta$ is assumed to be constant, we recover the $\beta$-plane approximation, usually exploited to modelize wave propagation in the equatorial area \cite{3,18,19,20,37,38}. Note that the ray Hamiltonian $\Omega^{(n)}$ for shallow water waves is exactly equal to $\omega_n$ in the $f$-plane approximation ($\beta = 0$) but in that case the tensor $F^{(n)}$ is not null.

\begin{figure}[h!]
    \begin{center}
    \includegraphics[scale=0.42]{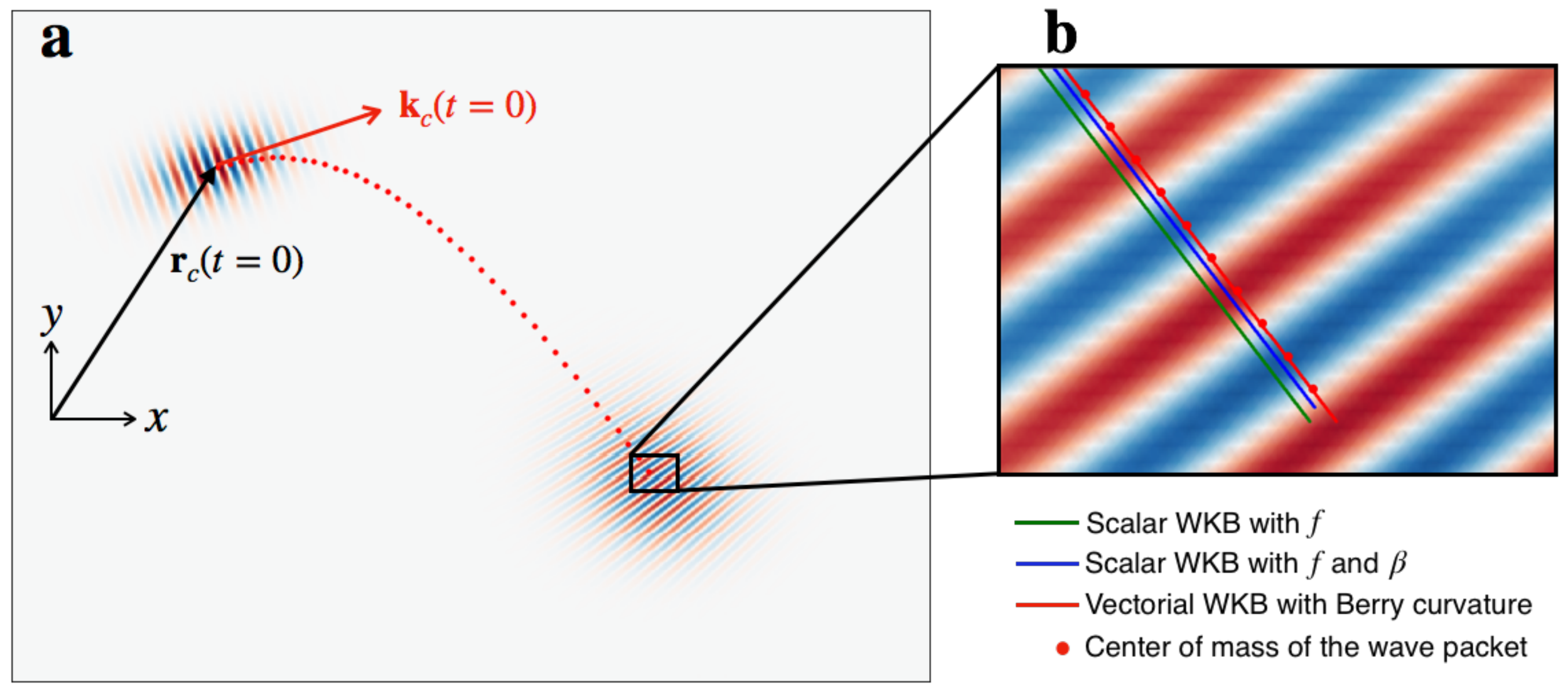}
    \end{center}
    \caption{\label{rayon} Simulation of an inertia-gravity wave packet in the northern hemisphere, computed with Dedalus \cite{40}. The initial wavelength is 1, $f_0 = 3$ the Coriolis parameter at middle ordinate of the frame, $\beta = 0.6$ and the simulation runs until $t = 32$. ($\bold{a}$) The red dots are the position of center of mass $\mathbf{r}_{c}(t)$ computed with the fields -- from definition \eqref{position-momentum} -- at regular steps of the simulation. ($\bold{b}$) We zoom around the last position of the packet and compare the real trajectory (red dots) with the ones predicted respectively by the geometric ray tracing equations \eqref{ray_bis} (red curve), scalar equations \eqref{ray_scalar} (blue curve) and elementary ray tracing equations \eqref{Hamilton} (green curve). Although the difference still appears small after $t = 32$, the geometric theory fits best the actual trajectory.}
\end{figure}

Just like traditional ray tracing methods (based on the eikonal equation \cite{17,38} or another scalar equation \cite{18,19,20,15,21}), we just need to know the properties of the homogeneous ($f$-plane) system to infer the leading-order deviations arising from inhomogeneities ($\beta$-plane). The upgrade brought by the variational method presented in section \ref{section:Berry} is that it adds the $f$-plane polarization relations (hence the Berry curvature) to the dispersion relation of the wave, thus accounting for its vectorial character. Then from the homogeneous model's eigenvalues $\omega_n$ and eigenmodes $U_n$, we just need to compute $\Omega^{(n)}$, defined by expression \eqref{correction}, and $F^{(n)}$, defined by expression \eqref{curvature_bis}, and inject them into equations \eqref{ray_tracing} to infer the trajectory of the corresponding wave packets (figure \ref{rayon}). We consider the shallow water model's different bands $n$: geostrophic waves $n = 0$ and Poincaré waves $n = \pm 1$. Since the shallow water equations are time-invariant, the ray Hamiltonian $\Omega^{(n)} (\mathbf{r}_{c}(t),\mathbf{k}_{c}(t))$ is a ray tracing constant of motion, which can be shown directly from the antisymmetry of equations \eqref{ray_tracing}.

\subsection{Berry curvature contribution to inertia-gravity waves eastward drift} \label{part:application}

Let us now compute the drift of shallow water vectorial wave packets induced by first order corrections to ray tracing equations, for the different bands $n = -1, 0, 1$ successively, by computing the observables $\Omega^{(n)}$ and $F^{(n)}$.

The connection ${\rm i} U_{0}^\dagger {\rm d} U_{0}$ of the geostrophic band $n = 0$ is null everywhere, hence so is its Berry curvature $F^{(0)}$ (see appendix \ref{appendix:eigenv}). Therefore, only the dynamical correction remains in the set of equations of motion \eqref{ray_tracing}. Expression \eqref{correction} with $n = 0$ yields
\begin{equation} \label{correction_geostrophic}
    \Omega^{(0)} (\mathbf{r}_{c},\mathbf{k}_{c}) = - \frac{\beta \, k_{xc}}{f(y_c)^2 + \mathbf{k}_{c}^2} \ ,
\end{equation}
which is nothing more than the dispersion relation of quasi-geostrophic planetary or Rossby waves \cite{15}. In that case the Berry curvature terms in equations \eqref{ray_tracing} drop and we simply retrieve Hamilton equations $\dot{\mathbf{r}}_{c} = \partial \Omega^{(0)} / \partial \mathbf{k}_{c}$ and $\dot{\mathbf{k}}_{c} = -\partial \Omega^{(0)} / \partial \mathbf{r}_{c}$, as if the equations of motion came from a scalar WKB ansatz: there is no geometrical effect for these wave rays, that is to say one scalar field -- for instance the geostrophic streamfunction -- contains all the information on the phase of the fields. Therefore, first order corrections to Rossby wave rays' dynamics only involve a corrected dynamical phase.

As for the higher-frequency Poincaré wave bands $n = \pm 1$, we can straightforwardly infer the properties of the one from the other: since the fields are real, a wave at $+\mathbf{k}$ on band $+1$ must behave as a wave at $-\mathbf{k}$ on band $-1$, which corresponds to the particle-hole-like symmetry $(\omega,\mathbf{k}) \rightarrow (-\omega,-\mathbf{k})$, because a real inertia-gravity wave packet is a combination of the two\footnote{This can be proven by injecting relation $U_{-1}[f,k_x,k_y] = -U_{+1}[-f,-k_x,-k_y]$ (see appendix \ref{appendix:eigenv}) into expressions \eqref{curvature_bis} and \eqref{correction}, and then the latter into ray tracing equations \eqref{ray_tracing}.}. Keeping that in mind we will work with $n = +1$ only. Expressions of $F^{(\pm 1)}$ and $\Omega^{(\pm 1)}$ are respectively given in appendices \ref{appendix:eigenv} and \ref{appendix:spacevaria}.

All terms combined, equations \eqref{ray_tracing} finally add up to
\begin{subequations} \label{ray_bis}
    \begin{align}
        \dot{x}_{c} &= \frac{k_{xc}}{\sqrt{f(y_c)^2 + \mathbf{k}_{c}^2}} \: + \: \frac{\beta(y_c)}{2 \left( f(y_c)^2 + \mathbf{k}_{c}^2 \right)} \label{ray_bis_x} \ , \\
        \dot{y}_{c} &= \frac{k_{yc}}{\sqrt{f(y_c)^2 + \mathbf{k}_{c}^2}} \label{ray_bis_y} \ , \\
        \dot{k}_{xc} &= 0 \ , \\
        \dot{k}_{yc} &= -\frac{\beta(y_c) \, f(y_c)}{\sqrt{f(y_c)^2 + \mathbf{k}_{c}^2}} \label{ray_bis_ky} \ .
    \end{align}
\end{subequations}
The third equation is a straightforward consequence of the invariance along the zonal direction whereas there is an effective force along the meridional direction. The only footprint of the medium's variation on the group velocity of inertia-gravity wave rays is along the zonal direction, while the geometrical corrections compensate the dynamical one for $\dot{y}_c$. On a clockwise rotating planet, up to the definition of the north pole, $\beta$ is positive everywhere and vanishes at the poles. Therefore only remains an eastward anomalous velocity along $x$ and a pseudo-force $\dot{\mathbf{k}}_c$ along $y$ in ray tracing equations \eqref{ray_bis}, both proportional to $\beta$, thus of order $\varepsilon$ in amplitude. There is an interesting parallel between this anomalous velocity under $\beta$-effect and the anomalous Hall effect \cite{28,34}: in this analogy the packet undergoes an electric field proportional to $\beta$ along the meridional direction $y$. As a consequence, an additional "anomalous" velocity arises, proportional and perpendicular to the electric field.

\subsection{Comparison with scalar ray tracing} \label{part:comparison}

As discussed in the introduction, the method used in \cite{18,19,20,15,21} to derive ray tracing equations for $v$ does not deal with the geometrical aspects of the problem as it ignores the polarization relations. We shall recall here the results obtained from this scalar framework and compare them to the ones we got with the vectorial ansatz and the variational principle exposed in section \ref{section:Berry}.

Following Bretherton \cite{17}, shallow water ray tracing equations in the presence of a varying Coriolis parameter are derived from the $f$-plane dispersion relation $\omega_{n}(\mathbf{r},\mathbf{k})$ given by expressions \eqref{spectrum}:
\begin{equation} \label{Hamilton}
    \dot{\mathbf{r}} = \frac{\partial \omega_n}{\partial \mathbf{k}} \ , \quad \text{and} \quad \dot{\mathbf{k}} = -\frac{\partial \omega_n}{\partial \mathbf{r}} \ .
\end{equation}
Ray coordinates are noted from now on $\mathbf{r} = (x,y)$ and $\mathbf{k} = (k_{x},k_{y})$ as well, without the index $c$, for simplification. Equations \eqref{Hamilton} yield a term of order $\beta$ for $\dot{\mathbf{k}}$ but no such correction in the expression of the group velocity $\dot{\mathbf{r}}$. 
In order to compute the $\beta$ correction to shallow water waves' group velocity, the traditional approach \cite{15,18,19,20,21} consists in applying Bretherton's method to the exact scalar equation for $v$ that includes the $\beta$-effect:
\begin{equation} \label{scalar_v}
    \partial_{ttt} v + f(y)^2 \partial_t v - \Delta \partial_t v - \beta(y) \partial_x v = 0 \ .
\end{equation}
The corrected dispersion relation is then obtained by formally replacing $\partial_t$ with $-{\rm i} \omega$ and $\boldsymbol{\nabla} = (\partial_x, \partial_y)$ with ${\rm i} \mathbf{k} = ({\rm i} k_{x}, {\rm i} k_{y})$ in \eqref{scalar_v}, which yields
\begin{equation} \label{dispersion}
    \omega^3 - \left( f(y)^2 + k_{x}^2 + k_{y}^2 \right) \omega - \beta(y) k_{x} = 0 \ .
\end{equation}

By differentiating equation \eqref{dispersion} with respect to $\mathbf{k}$, one eventually gets the following expression for the group velocity:
\begin{equation} \label{scalar_velocity}
    \frac{\partial \omega}{\partial \mathbf{k}} = \frac{\mathbf{k} + \frac{\beta}{2 \omega} \hat{\mathbf{e}}_{x}}{\omega + \frac{\beta k_{x}}{2 \omega^2}} \ ,
\end{equation}
where $\hat{\mathbf{e}}_{x}$ is the unitary basis vector along the local zonal direction, pointing eastward, and $\omega$ is one of the solutions of the dispersion relation \eqref{dispersion}. This result can be found in Vallis' book \cite{15} (p. 311). It echoes directly the scalar WKB solution of equation \eqref{scalar_v}. However using only this equation for ray tracing is problematic whenever $\beta \neq 0$, because the other fields of the problem satisfy different equations:
\begin{subequations} \label{scalar_u-n}
    \begin{align}
        \partial_{ttt} u + f(y)^2 \partial_t u - \Delta \partial_t u + \beta(y) \partial_x u &= -2 \beta(y) \partial_y v - \beta'(y) v \label{scalar_u} \ , \\
        \partial_{ttt} \eta + f(y)^2 \partial_t \eta - \Delta \partial_t \eta + \beta(y) \partial_x \eta &= 2 \beta(y) f(y) v \label{scalar_n} \ .
    \end{align}
\end{subequations}
This multi-component aspect leads to inconsistent WKB ansätze for the different scalar fields of the problem \cite{22}. Yet accounting for the three fields is crucial for computing ray observables such as the ones defined by relations \eqref{energy} or \eqref{position-momentum}, and therefrom to properly derive ray tracing equations.

When $\beta \sim \varepsilon$, both scalar and vectorial methods yield the same leading order ray tracing equation through the $f$-plane dispersion relation, whose solutions are given by expressions \eqref{spectrum}. The $\beta$ term in the LHS of equation \eqref{scalar_v} appears only at next order. We explained that $\beta$ also induces additional geometric corrections to ray tracing equations, owing to the multi-component nature of the problem. The scalar approach is therefore inconsistent at order $\varepsilon$. 

\begin{figure}[h!]
    \begin{center}
    \includegraphics[scale=0.4]{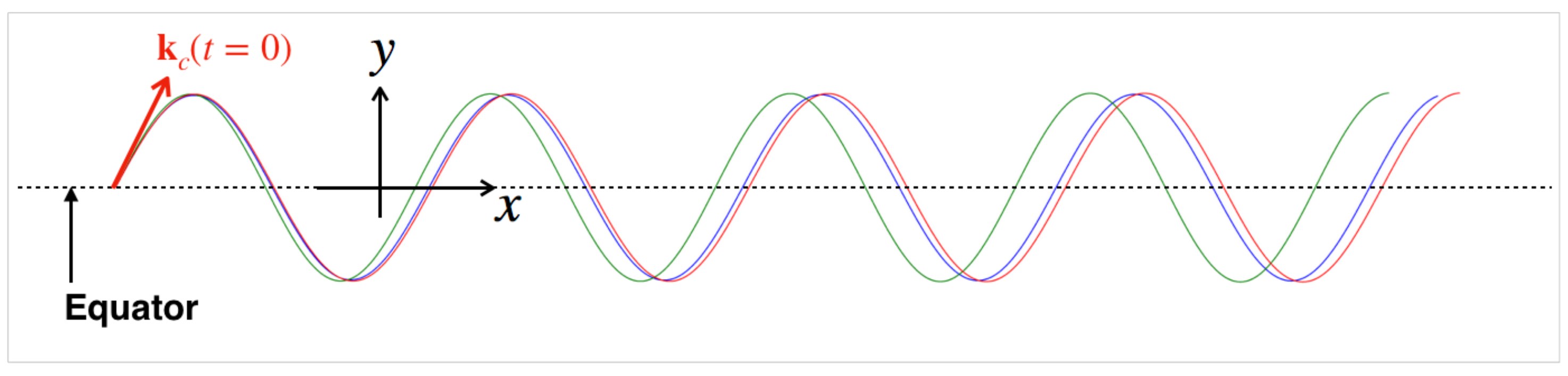}
    \end{center}
    \caption{\label{equatorial} Ray trajectory of an inertia-gravity wave packet oscillating eastward around the equator, predicted by the different theories (same color code as in figure \ref{rayon}). The initial wavelength is 2.8, $\beta = 0.25$ and the ray trajectories are plotted to $t = 240$. The vectorial theory anticipates a packet heading east to be faster than within the scalar framework. Conversely, a packet heading west would be slower.}
\end{figure}

Thus the only way to properly take into account the inhomogeneity parameter $\beta$ is to introduce it as a perturbation in the vectorial problem. Keeping that in mind, it is instructive to compare the group velocity \eqref{scalar_velocity} derived within the traditional scalar WKB  framework to the group velocity \eqref{ray_bis} derived within our vectorial framework. By differentiating expression \eqref{dispersion} with respect to $\mathbf{k}$ and $\mathbf{r}$ and expanding the result at order one in $\beta \sim \varepsilon$, one gets the following ray tracing equations for Poincaré wave rays\footnote{Again we work with the positive solution ($n = +1$), which reads $\omega(\mathbf{r},\mathbf{k}) = \sqrt{f(y)^2 + \mathbf{k}^2} + \frac{\beta k_x}{2 \left( f(y)^2 + \mathbf{k}^2 \right)} + \mathcal{O}\left( \varepsilon^2 \right)$.}:
\begin{subequations} \label{ray_scalar}
    \begin{align}
        \dot{\mathbf{r}} &= \frac{\partial \omega}{\partial \mathbf{k}} = \frac{\mathbf{k}}{\sqrt{f^2 + \mathbf{k}^2}} + \beta \left[ \frac{\hat{\mathbf{e}}_{x}}{2 \left( f^2 + \mathbf{k}^2 \right)} - \frac{k_{x} \mathbf{k}}{\left( f^2 + \mathbf{k}^2 \right)^2} \right] \label{ray_scalar_r} \ , \\
        \dot{\mathbf{k}} &= -\frac{\partial \omega}{\partial \mathbf{r}} = -\beta \frac{f \hat{\mathbf{e}}_{y}}{\sqrt{f^2 + \mathbf{k}^2}} \label{ray_scalar_k} \ .
    \end{align}
\end{subequations}
The latter must be compared to expressions \eqref{ray_bis_x} and \eqref{ray_bis_y} obtained with the vectorial ansatz. The most striking difference is that the correction proportional to $\beta$ in the group velocity \eqref{ray_scalar_r} is not always pointing eastward, contrary to what is predicted by the vectorial method. Accounting for the geometrical character of the inertia-gravity wave rays thus adds a systematic eastward term to its group velocity compared to the expected value of the scalar WKB approximation, as illustrated in figure \ref{equatorial}. Figure \ref{equatorial} presents the case of equatorial oscillation \cite{15,19,21}, an instructive example showing the enhanced eastward group velocity of an inertia-gravity wave packet, predicted by accounting for the Berry curvature of the Poincaré band in ray tracing equations. However one must keep in mind that in the equatorial region -- as opposed to mid-latitude ray tracing (figure \ref{rayon}) -- below a typical latitude called the equatorial radius of deformation ($L_{eq} = \sqrt{c / \beta_{eq}}$, for the first baroclinic mode in the equatorial ocean on Earth we have $c \simeq 2$ m.s$^{-1}$, thus $L_{eq} \simeq 300$ km \cite{15}), our scaling fails down: on the one hand the Coriolis parameter vanishes at the equator, thus so does the separation in frequency between inertia-gravity modes and planetary modes, that can hybridize. And on the other hand the wave packet is subjected to strong dispersion in the equatorial region, that eventually spreads its envelope to a scale comparable with $L$.

One can wonder what are the typical orders of magnitude for this vectorial correction in dimensional units, say for oceanic inertia-gravity waves on Earth, with $c = 2$ m.s$^{-1}$ and $\beta = 2.3 \: . 10^{-11}$ m$^{-1}$.s$^{-1}$ (in the following all quantities are given in their natural units). Subtracting the projection along $x$ of equation \eqref{ray_scalar_r} to expression \eqref{ray_bis_x}, the zonal projection of the difference of group velocity between the vectorial and the scalar method reads
\begin{equation} \label{zonal_correction}
    \Delta v_x = \frac{\beta k_{x}^2}{\left(\left( f/c \right)^2 + \mathbf{k}^2 \right)^2} \sim \beta \lambda^2 \sim c \left( \frac{\lambda}{L_{eq}} \right)^2 \ ,
\end{equation}
which is a constant of motion with both methods, and shall be compared to the leading order zonal velocity $c k_x / \sqrt{\left( f/c \right)^2 + \mathbf{k}^2}$. For inertia-gravity waves of wavelength smaller than the equatorial radius of deformation ($\lambda \sim 100$ km), relation \eqref{zonal_correction} adds up to a vectorial correction of about one to ten percent of $c$, typically, to the zonal group velocity, thus yielding an overall zonal shift of order $\lambda$ after a time $t = \lambda / \Delta v_x \sim (\beta \lambda)^{-1}$, that is typically a month. The corresponding total zonal displacement of the wave packet after such time if about $c/(\beta \lambda) \sim 6000$ km.

\section{Conclusion} \label{section:conclusion}

Using the classical example of the linear $\beta$-plane shallow water model, we showed how the vectorial nature of eigenmodes is involved their ray dynamics, both through the $\beta$-effect and a geometrical observable: the Berry curvature. In a way that could not be derived from a scalar approach, these joint contributions yield an eastward deviation of Poincaré wave rays' trajectory, exhibiting a geophysical analog with the anomalous Hall effect. The ray tracing equations that we derive are based on coherent hypotheses with respect to the scales characterizing the wave packet and the variations of the Coriolis parameter, in contrast with scalar derivations of equatorial ray tracing equations that can be found in classical textbooks.

Previous works had revealed geometric phase effects in geophysical waves \cite{25,22,27}, although it is rarely mentioned in geophysical fluid dynamics textbooks, with the notable exception of \cite{16}. Our contribution here has been to exhibit a quantitative manifestation of the Berry curvature in ray tracing equations, with application to shallow water waves. This manifestation is yet subtle in a practical point of view. Indeed the Berry curvature of inertia-gravity wave bands is highest at small wave vectors, thus for large wavelength (figure \ref{courbure}). Yet our analysis is based on a scaling that assumes sufficiently small wavelength. Furthermore, large Berry curvature is concomitant to large dispersion as both effects are associated with the proximity of a degeneracy point between wave bands. In practice, large dispersion tends to spread the envelop of the wavepacket. This effect of dispersion, together with dissipation and nonlinearities, will need to be addressed in future works.

Our approach thus highlights the multi-component nature of geophysical wave dynamics, where the primitive set of equations always involve the time derivative of several fields coupled together. The dynamics is sometimes reduced down to a scalar equation, which can conveniently be expressed in a Lagrangian framework different than the one used in this paper \cite{41}. These scalar field Lagrangian approaches are very useful as they allow one to describe nonlinear effects. However, information of the underlying multi-component dynamics is not encoded into such scalar Lagrangians. This is why we used here an alternative variational approach that takes into account the multi-component nature of the wave equations. It cannot account for nonlinear effects, but it is suitable to capture Berry curvature effects on ray trajectories.

Since Berry curvature most often comes along with multi-component wave problems in the presence of broken discrete symmetries, it is natural to expect its manifestation on a wider class of systems, from small scales hydrodynamics to large scales geophysics and astrophysics, as well as in the domain of seismic and elastic waves.

\appendix
\appendixpage

\section{eigenmodes and Berry curvature of the $f$-plane shallow water system} \label{appendix:eigenv}

We give here an expression of the eigenmodes and Berry curvature of the $f$-plane shallow water model for both geostrophic ($n = 0$) and inertia-gravity ($n = \pm 1$) wave bands, which are purely geometrical features of the eigenspaces supporting the vectorial plane wave solutions of the $f$-plane problem.

As exposed in section \ref{section:shallow-water}, looking for dispersion and polarization relations of the Fourier solutions of the $f$-plane shallow water model amouts to solving the eigenvalue equation \eqref{symbol}. For a given value of $(f,k_x,k_y)$ the eigenvalues are given by expressions \eqref{spectrum}, and the corresponding $\mathbb{C}^3$ basis of normalized eigenmodes are
\begin{equation} \label{U_0}
    U_{0}[f,k_x,k_y] = \frac{1}{\sqrt{f^2 + k^2}} \begin{pmatrix} k_y \\ -k_x \\ {\rm i} f \\ \end{pmatrix} = \begin{pmatrix} \sin{\theta} \sin{\phi} \\ -\sin{\theta} \cos{\phi} \\ {\rm i} \cos{\theta} \\ \end{pmatrix}
\end{equation}
for the geostrophic band, and
\begin{equation} \label{U_pm}
    U_{\pm 1}[f,k_x,k_y] = \frac{1}{k \sqrt{2}} \begin{pmatrix} k_x \pm {\rm i} \frac{f k_y}{\sqrt{f^2 + k^2}} \\ k_y \mp {\rm i} \frac{f k_x}{\sqrt{f^2 + k^2}} \\ \pm \frac{k^2}{\sqrt{f^2 + k^2}} \\ \end{pmatrix} = \frac{1}{\sqrt{2}} \begin{pmatrix} \cos{\phi} \pm {\rm i} \cos{\theta} \sin{\phi} \\ \sin{\phi} \mp {\rm i} \cos{\theta} \cos{\phi} \\ \pm \sin{\theta} \\ \end{pmatrix}
\end{equation}
for inertia-gravity bands, all three being defined up to a gauge phase. Due to the linear property $H[f,k_x,k_y] = -H[-f,-k_x,-k_y]$ we have $U_{-1}[f,k_x,k_y] = -U_{+1}[-f,-k_x,-k_y]$, again up to a choice of gauge. In the previous we have noted $k \equiv \left\Vert\mathbf{k}\right\Vert = \sqrt{k_{x}^{2} + k_{y}^{2}}$ and $(\phi,\theta)$ the spherical coordinates of the $\mathbb{R}^3$ vector $\begin{pmatrix} k_x \: k_y \: f \end{pmatrix}^t$ of norm $\sqrt{f^2 + k^2}$. From expressions \eqref{U_pm} we arrive at
\begin{equation} \label{connections}
    {\rm i} U_{0}^\dagger {\rm d} U_{0} = 0 \ , \quad {\rm i} U_{\pm 1}^\dagger {\rm d} U_{\pm 1} = \mp \cos{\theta} {\rm d} \phi
\end{equation}
for the connections of the eigenbundles, hence
\begin{equation} \label{curvatures}
    \mathcal{F}^{(0)} = 0 \ , \quad \mathcal{F}^{(\pm 1)} = \pm \sin{\theta} {\rm d} \theta \wedge {\rm d} \phi
\end{equation}
for the Berry curvatures. Keeping in mind that in the three-dimensional case the coefficients of $\mathcal{F}^{(n)}$ are given by the curl of ${\rm i} U_{n}^\dagger \boldsymbol{\nabla} U_{n}$, which is equal to $\mp \frac{\cot{\theta}}{\sqrt{f^2 + k^2}} \hat{\mathbf{e}}_{\phi}$ if $n = \pm 1$, we have for the latter
\begin{equation} \label{curl}
    \begin{pmatrix} \mathcal{F}^{(\pm 1)}_{k_y f} \\ \mathcal{F}^{(\pm 1)}_{f k_x} \\ \mathcal{F}^{(\pm 1)}_{k_x k_y} \\ \end{pmatrix} = \boldsymbol{\nabla} \times \left( {\rm i} U_{\pm 1}^\dagger \boldsymbol{\nabla} U_{\pm 1} \right) = \pm \frac{1}{\sqrt{f^2 + k^2}^{\: 3}} \begin{pmatrix} k_x \\ k_y \\ f \\ \end{pmatrix} \ ,
\end{equation}
in the natural cartesian basis of $\begin{pmatrix} k_x \: k_y \: f \end{pmatrix}^t$. Finally, formally replacing the variable $f$ in the previous, thanks to ${\rm d} f = \beta \, {\rm d} y$, yields the expressions of the only non-zero coefficients\footnote{Since the Coriolis parameter has no dependence in the longitudinal direction $x$ and $F^{(n)}$ are antisymmetric tensors, the Berry curvature is entirely defined by the 3 coefficients $F^{(\pm 1)}_{y k_{x}}$, $F^{(\pm 1)}_{y k_{y}}$ and $F^{(\pm 1)}_{k_{x} k_{y}}$. } of the Berry curvature $F^{(\pm 1)}$, used to derive ray tracing equations \eqref{ray_bis}:
\begin{subequations} \label{Berry_inertia}
    \begin{align}
        F^{(\pm 1)}_{k_{y} y} (\mathbf{r},\mathbf{k}) &= -F^{(\pm 1)}_{y k_{y}} (\mathbf{r},\mathbf{k}) = \pm \, \frac{\beta \, k_{x}}{\sqrt{f(y)^2 + \mathbf{k}^2}^{\: 3}} \ , \\
        F^{(\pm 1)}_{y k_{x}} (\mathbf{r},\mathbf{k}) &= -F^{(\pm 1)}_{k_{x} y} (\mathbf{r},\mathbf{k}) = \pm \, \frac{\beta \, k_{y}}{\sqrt{f(y)^2 + \mathbf{k}^2}^{\: 3}} \ , \\
        F^{(\pm 1)}_{k_{x} k_{y}} (\mathbf{r},\mathbf{k}) &= -F^{(\pm 1)}_{k_{y} k_{x}} (\mathbf{r},\mathbf{k}) = \pm \, \frac{f(y)}{\sqrt{f(y)^2 + \mathbf{k}^2}^{\: 3}} \ ,
    \end{align}
\end{subequations}

\section{Variational derivation of ray tracing equations} \label{appendix:spacevaria}

In this appendix we derive the ray equations for ray tracing starting from ansatz \eqref{space-ansatz} for the wave packet. We pay particular attention to the necessary approximations to get the expansion of the Lagrangian \eqref{Lagrangian_red}.

\subsection{Reduction of the Lagrangian $\mathcal{L'}$} \label{part:lagrangian}

Firstly definition \eqref{Lagrangian} can be separated as
\begin{equation} \label{Lagrangian-parts}
    \mathcal{L'} = \frac{\langle \psi' | {\rm i} \partial_t | \psi' \rangle}{\langle \psi' | \psi' \rangle} - \frac{\langle \psi' | \hat{\mathcal{H}} | \psi' \rangle}{\langle \psi' | \psi' \rangle} \ .
\end{equation}
The second term, defined in relation \eqref{ray_Hamiltonian}, will be treated separately. Using the ansatz \eqref{space-ansatz}, the first term yields
\begin{equation} \label{part1}
    \frac{\langle \psi' | {\rm i} \partial_t | \psi' \rangle}{\langle \psi' | \psi' \rangle} = \frac{\int A^2 \left( -\partial_t \Phi + {\rm i} U_{n}^\dagger \partial_t U_{n} \right)}{\int A^2} + \frac{{\rm i}}{2} \frac{{\rm d}}{{\rm d} t} \ln{\left( \int A^2 \right)} \ ,
\end{equation}
with the notation $\int A^2 \equiv \int {\rm d}^2 r A(\mathbf{r},t)^2$. To lighten presentation, variables of functions have been omitted, i.e. $\Phi$ stands for $\Phi(\mathbf{r},t)$ and $U_{n}$ for $U_{n}[f(\mathbf{r}),\mathbf{k}_c (t)]$, for instance. Since the operator $\hat{\mathcal{H}}$ is Hermitian, the norm of $| \psi \rangle$ is left constant by the dynamics\footnote{Indeed $\frac{{\rm i}}{2} \frac{{\rm d}}{{\rm d} t} \ln{\left( \int A^2 \right)} = {\rm i} \frac{{\rm d}}{{\rm d} t} \left\Vert\psi'\right\Vert$ and we can restrain the search of a solution to the space of functions of norm constant with respect to time, because $\hat{\mathcal{H}}$ is Hermitian: that way the Lagrangian $\mathcal{L'}$ is real. One would directly remove the second term in equation \eqref{part1} from the variational problem because it is an exact time derivative, but if the Lagrangian were not real the Schrödinger-like equation \eqref{schrödinger} would not be equivalent to the corresponding variational problem \cite{12}.}, therefore the second term in equation \eqref{part1} is null, leaving us only with the first term, which is real. Similarly, the expression of the mean wave vector defined in equations \eqref{position-momentum} is simplified into
\begin{equation} \label{momentum_bis}
    \mathbf{k}_{c}(t) = \frac{\int A^2 \left( \boldsymbol{\nabla} \Phi - {\rm i} U_{n}^\dagger \boldsymbol{\nabla} U_{n} \right)}{\int A^2} \ .
\end{equation}
Let us now simplify expressions \eqref{part1} and \eqref{momentum_bis} by approximating the integrals weighted by $A^2$. The Taylor expansion of a general function $g(\mathbf{r},t)$ around $\mathbf{r}_{c}(t)$ reads
\begin{equation} \label{Taylor}
    g(\mathbf{r},t) = g(\mathbf{r}_{c},t) + \left( \mathbf{r} - \mathbf{r}_{c} \right) \cdot \nabla g(\mathbf{r}_{c},t) + \mathcal{O}\left(\left( \mathbf{r} - \mathbf{r}_{c} \right)^2 \varepsilon^{-2} g_0 \right) \ ,
\end{equation}
if $\varepsilon^{-1}$ is the variation scale of the function $g$ of order $g_0$. By performing the integral of the previous weighted by $A^2$, the second term in the RHS of equation \eqref{Taylor} yields 0, by definition of $\mathbf{r}_c$, whereas by definition of $\alpha$ the third one remains and we get:
\begin{equation} \label{part2}
    \frac{\int A^2 g(\mathbf{r},t)}{\int A^2} = g(\mathbf{r}_{c},t) + \mathcal{O}\left( \left( \frac{\varepsilon}{\alpha} \right)^2 g_0 \right) \ .
\end{equation}
The derivatives of the slowly varying functions in integrals \eqref{part1} and \eqref{momentum_bis} reveal terms of order 1, with $-\partial_t \Phi_0 / \varepsilon = -\partial_T \Phi_0$ and $\nabla \Phi_0 / \varepsilon = \nabla_\mathbf{R} \Phi_0$, and $\varepsilon$ with $\varepsilon \left( -\partial_T \Phi_1 + {\rm i} U_{n}^\dagger \partial_T U_{n} \right)$ and $\varepsilon \left( \nabla_\mathbf{R} \Phi_1 - {\rm i} U_{n}^\dagger \nabla_\mathbf{R} U_{n} \right)$. Therefore we have here $g_0$ of order 1 at most and we arrive at
\begin{equation} \label{part3}
    \frac{\langle \psi' | {\rm i} \partial_t | \psi' \rangle}{\langle \psi' | \psi' \rangle} = -\partial_t \Phi (\mathbf{r}_{c},t) + {\rm i} U_{n} [f(y_c),\mathbf{k}_c (t)]^\dagger \partial_t U_{n} [f(y_c),\mathbf{k}_c (t)] + \mathcal{O}\left(\left( \frac{\varepsilon}{\alpha} \right)^2\right) \ ,
\end{equation}
and
\begin{equation} \label{momentum_bbis}
    \mathbf{k}_c = \nabla \Phi (\mathbf{r}_{c},t) - {\rm i} U_{n} [f(y_c),\mathbf{k}_c (t)]^\dagger \nabla U_{n} [f(y_c),\mathbf{k}_c (t)] + \mathcal{O}\left(\left( \frac{\varepsilon}{\alpha} \right)^2\right) \ .
\end{equation}
This last expression is the same as \eqref{k=K}, only here for the purpose of the following development we do not need to separate the terms of order 1 and $\varepsilon$. Noticing now that $\frac{{\rm d}}{{\rm d} t} \phi (\mathbf{r}_{c}(t),t)$ is equal to $\left( \partial_t + \dot{\mathbf{r}}_{c} \cdot \nabla \right) \phi (\mathbf{r}_{c},t)$ for any function $\phi (\mathbf{r},t)$, we can combine equation \eqref{part3} with equation \eqref{momentum_bbis} and expressions \eqref{ray_Hamiltonian}, and the Lagrangian reads
\begin{equation} \label{part4}
    \mathcal{L'} + \frac{{\rm d}}{{\rm d} t} \Phi (\mathbf{r}_{c}(t),t) = \dot{\mathbf{r}}_{c} \cdot \mathbf{k}_{c} - \Omega^{(n)}(\mathbf{r}_{c},\mathbf{k}_{c}) + {\rm i} U_{n}(\mathbf{r}_{c},\mathbf{k}_{c})^\dagger \frac{{\rm d}}{{\rm d} t} U_{n}(\mathbf{r}_{c},\mathbf{k}_{c}) + \mathcal{O}\left(\left( \frac{\varepsilon}{\alpha} \right)^2\right) \ .
\end{equation}
This last relation corresponds to expression \eqref{Lagrangian_red} with the notation $\chi(t) \equiv \Phi (\mathbf{r}_{c}(t),t)$.

\subsection{Expression of the ray Hamiltonian} \label{part:hamiltonian}

Let us now prove expression \eqref{correction} for $\Omega^{(n)}$ in the case of the shallow water model. First of all we notice that the Hamiltonian operator defined in relations \eqref{operator} reads $\hat{\mathcal{H}}(\mathbf{r},\boldsymbol{\nabla}) = f(\mathbf{r}) \frac{\partial H}{\partial f} - {\rm i} \frac{\partial H}{\partial \mathbf{k}} \cdot \boldsymbol{\nabla}$, using the formal derivatives of the symbol $H$ defined in relation \eqref{symbol}. In addition one can check that $U_{n}^\dagger \frac{\partial H}{\partial \mathbf{k}} U_{n}$ is equal to $\mathbf{c}_n \equiv \partial \omega_{n} / \partial \mathbf{k}$, the group velocity of the $n^{th}$ band of the $f$-plane system. Then using ansatz \eqref{space-ansatz} yields
\begin{equation} \label{Omega1}
    \langle \hat{\mathcal{H}} \rangle \equiv \frac{\langle \psi | \hat{\mathcal{H}} | \psi \rangle}{\langle \psi | \psi \rangle} = \frac{\int A^2 \left[ \mathbf{c}_n \cdot \boldsymbol{\nabla} \Phi + f U_{n}^\dagger \frac{\partial H}{\partial f} U_{n} -{\rm i} U_{n}^\dagger \frac{\partial H}{\partial \mathbf{k}} \cdot \boldsymbol{\nabla} U_{n} \right] - {\rm i} \mathbf{c}_n \cdot \boldsymbol{\nabla} \left( \frac{A^2}{2} \right)}{\int A^2} \ .
\end{equation}
Integrating by parts the last term in the integral of expression \eqref{Omega1}, and noticing that the symbol reads $H[f(\mathbf{r}),\mathbf{k}_{c}] = f(\mathbf{r}) \frac{\partial H}{\partial f} + \frac{\partial H}{\partial \mathbf{k}} \cdot \mathbf{k}_{c}$, equation \eqref{Omega1} yields
\begin{equation} \label{Omega2}
    \langle \hat{\mathcal{H}} \rangle = \frac{\int A^2 \left[ \omega_{n} + \mathbf{c}_n \cdot \left( \boldsymbol{\nabla} \Phi -\mathbf{k}_{c} \right) +\frac{{\rm i}}{2} \boldsymbol{\nabla} \cdot \mathbf{c}_n -{\rm i} U_{n}^\dagger \frac{\partial H}{\partial \mathbf{k}} \cdot \boldsymbol{\nabla} U_{n} \right]}{\int A^2} \ .
\end{equation}
Besides, knowing that $\sum_{m \in \{-1,0,+1\}} U_{m} U_{m}^\dagger = I_3$ (the 3 by 3 identity matrix) and that the connection ${\rm i} U_{n}^\dagger \nabla U_{n}$ is real, one gets
\begin{equation} \label{Omega3}
        \frac{{\rm i}}{2} \boldsymbol{\nabla} \cdot \mathbf{c}_n -{\rm i} U_{n}^\dagger \frac{\partial H}{\partial \mathbf{k}} \cdot \boldsymbol{\nabla} U_{n} = -\mathbf{c}_n \cdot {\rm i} U_{n}^\dagger \boldsymbol{\nabla} U_{n} + \mathcal{I}m \left( \sum_{m \neq n} U_{n}^\dagger \frac{\partial H}{\partial \mathbf{k}} U_{m} \cdot U_{m}^\dagger \boldsymbol{\nabla} U_{n} \right) \ .
\end{equation}
For $m \neq n$, one can prove that $U_{m}^\dagger \nabla U_{n}$ is equal to $\frac{U_{m}^\dagger \nabla H U_{n}}{\omega_{n} - \omega_{m}}$ \cite{1}, therefore relation \eqref{Omega2} yields
\begin{equation} \label{Omega4}
    \langle \hat{\mathcal{H}} \rangle = \frac{\int A^2 \left[ \omega_{n} + \mathbf{c}_n \cdot \left( \boldsymbol{\nabla} \Phi -\mathbf{k}_{c} - {\rm i} U_{n}^\dagger \boldsymbol{\nabla} U_{n} \right) - \mathcal{I}m \left( \sum_{m \neq n} \frac{U_{n}^\dagger \frac{\partial H}{\partial \mathbf{r}} U_{m} \cdot U_{m}^\dagger \frac{\partial H}{\partial \mathbf{k}} U_{n}}{\omega_{n} - \omega_{m}} \right) \right]}{\int A^2} \ .
\end{equation}
With the approximation of integrals weighted by $A^2$ explained in part \ref{part:lagrangian} of this appendix, the second term in expression \eqref{Omega4} being straightforward to deal with via relation \eqref{momentum_bbis} and the third one of order $\varepsilon$, equation \eqref{Omega4} yields
\begin{equation} \label{Omega5}
    \langle \hat{\mathcal{H}} \rangle = \omega_{n} - \mathcal{I}m \left( \sum_{m \neq n} \frac{U_{n}^\dagger \frac{\partial H}{\partial \mathbf{r}_c} U_{m} \cdot U_{m}^\dagger \frac{\partial H}{\partial \mathbf{k}_c} U_{n}}{\omega_{n} - \omega_{m}} \right) + \mathcal{O}\left(\left( \frac{\varepsilon}{\alpha} \right)^2\right) \equiv \Omega^{(n)} + \mathcal{O}\left(\left( \frac{\varepsilon}{\alpha} \right)^2\right) \ .
\end{equation}
An equivalent writing of $\Omega^{(n)}$, as in \cite{7}, is given by
\begin{equation} \label{Omega5b}
    \Omega^{(n)} = \omega_{n} - \mathcal{I}m \left[ \frac{\partial U_{n}^{\dagger}}{\partial r_{\mu c}}(\omega_{n} I_3 - H)\frac{\partial U_{n}}{\partial k_{\mu c}} \right] \ .
\end{equation}
All functions in expressions \eqref{Omega5b} are valued at $(\mathbf{r}_{c},\mathbf{k}_{c})$, which demonstrates that $\Omega^{(n)}$ is indeed a function of $\mathbf{r}_c$ and $\mathbf{k}_c$ only, at leading order in the gradient corrections. This proves relation \eqref{correction}. Using expression \eqref{Omega5} for Rossby waves ($n = 0$) one gets relation \eqref{correction_geostrophic} for $\Omega$. For Poincaré waves ($n = \pm 1$), it yields:
\begin{equation} \label{correction_inertia}
    \Omega^{(\pm 1)} (\mathbf{r}_{c},\mathbf{k}_{c}) = \pm \, \sqrt{f(y_c)^2 + \mathbf{k}_{c}^2} \: - \: \frac{\beta \,  k_{xc}}{2 \left( f(y_c)^2 + \mathbf{k}_{c}^2 \right)} \ .
\end{equation}

\enlargethispage{20pt}

\dataccess{We declare that all information required to reproduce the paper is held within the paper.}

\aucontribute{N.P. conducted research and performed the numerical calculations. P.D. and A.V. supervised the study. All authors discussed the physics, participated in writing and revising the manuscript, gave final approval for publication and agree to be held accountable for the work performed therein.}

\competing{The authors declare no competing financial interest.}

\funding{This project was supported by the national grant ANR-18-CE30-0002-01. N.P. was funded by a PhD grant allocation Contrat doctoral Normalien.}

\end{document}